\documentclass[doublecol]{epl2} 
\usepackage{amssymb}

\title{Bose-Einstein condensation on a superconducting atom chip}
\shorttitle{BEC on a superconducting atom chip} 

\author{C. Roux\inst{1} \and A. Emmert\inst{1} \and A. Lupascu\inst{1} \and T.
Nirrengarten\inst{1} \and G. Nogues\inst{1}\thanks{E-mail: \email{nogues@lkb.ens.fr}} \and M. Brune\inst{1} \and J.-M.
Raimond\inst{1} \and S. Haroche\inst{1,2}}
\shortauthor{C. Roux \etal}

\institute{                    
  \inst{1} Laboratoire Kastler Brossel, ENS, UPMC, CNRS - 24 rue Lhomond, 75005 Paris, France\\
  \inst{2} Coll\`{e}ge de France - 11 place Marcelin Berthelot, 75005 Paris,
France
}

\abstract{We have produced a Bose-Einstein condensate (BEC) on an atom
chip using only
superconducting wires in a cryogenic environment. We observe the onset of
condensation for $1\cdot 10^4$ atoms at a temperature of
100~nK. This result opens
the way for
studies of atom losses and decoherence in a BEC interacting with a
superconducting surface. Studies of dipole-blockade with long-lived Rydberg atoms in a small and dense atomic
sample are underway.
}

\begin{document}

\maketitle

Atom chips allow to trap ultracold atoms in the vicinity of micron-sized current
carrying wires~\cite{TR_ZIMMERMANRMPCHIP07} or permanent magnetic
structures~\cite{TR_HINDSREVIEWFERROMAG99}. Microfabrication techniques
offer the opportunity to engineer a wide range of magnetic potentials~\cite{TR_HANSCHCONVEYOR01,TR_PRENTISSBEAMSPLITTERCHIP05}. Moreover, atom chips provide compact traps which could have important applications in metrology experiments based on atom
interferometers~\cite{TR_REICHELCOHERENCECHIPCLOCK04,TR_SCHMIEDMAYERDBLWELL05,TR_ZIMMERMANBLOCH05,TR_KETTERLECHIPSQUEEZ07}.

Atomic samples prepared in a degenerate
bosonic~\cite{MX_HANSCHBECCHIP01,TR_ZIMMERMANNBECCHIP01} or
fermionic~\cite{TR_THYWISSENFERMIONCHIP06} collective quantum state have in this
context a particular interest. An essential requirement is however to obtain, in
close vicinity of the chip, a long enough lifetime for the condensed atomic
sample. The lifetime over normal metal chips at room temperature is significantly affected by thermal current fluctuations (Nyquist noise) in the conducting surfaces. The resulting random magnetic field induces transitions towards untrapped magnetic states~\cite{TR_HENKELMETALNOISE05}. 
A possible solution to this problem is to produce the
Bose-Einstein condensate (BEC) over permanent non-conducting
structures~\cite{TR_HINDSBECVIDEOTAPE05} or dielectric
surfaces~\cite{TR_VULETICCHIPNOISE04}. Another promising possibility is to operate at low temperatures in the vicinity
of superconductors. The near-field magnetic noise spectrum above a superconducting
slab is expected to be significantly different from that of a metal at room temperature~\cite{TR_REKDALSUPERCONDNOISE06,TR_HINDSELIASHBERG07}. 

In addition, superconducting atom chips pave the way to the coupling of atomic samples to mesoscopic superconducting devices. A superposition of flux states in a SQUID could affect the BEC trapping potential, leading to macroscopic entanglement~\cite{TR_SINGHBECLOOP07}. One can achieve even stronger coupling if the sample is excited towards Rydberg states~\cite{QI_LUKINTWORYDBERG04} and electrically coupled to the field of a strip line cavity~\cite{QI_DEMILLEZOLLERPOLARMOLECULES}.

These perspectives have triggered efforts to develop
cryogenic atom chips made of superconducting
materials~\cite{ENS_CHIPSUPRA06,TR_SHIMIZUPERMANENTCHIP07}. Very promising
trapping lifetimes have already been observed~\cite{ENS_CHIPSUPRA06}. We report here on the first observation of a BEC on a superconducting niobium
atom chip.
Bose-Einstein condensates with $1\cdot 10^4$
atoms and temperatures below 100~nK are produced at a distance of 85~$\mu$m from
the chip
surface. They can be compressed and brought as close as 50~$\mu$m from the chip.
The cloud is then contained in a volume of a few $\mu$m$^3$. This opens the way
to systematic studies of the cloud lifetime in the vicinity of a type-II
superconductor. Moreover, the achieved atomic
densities are compatible with the observation of strong dipole-blockade if the
cloud is excited towards Rydberg
states~\cite{TR_EYLERBLOCKADE04,TR_WEIDEMULLERDENSE04,TR_PILLETBLOCKADE06,
TR_PFAUBLOCKADE07}. 

\section{Experimental setup and loading of the atom-chip trap}

\begin{figure}[h]
(a)\\
\centerline{\includegraphics[width=5.5cm]{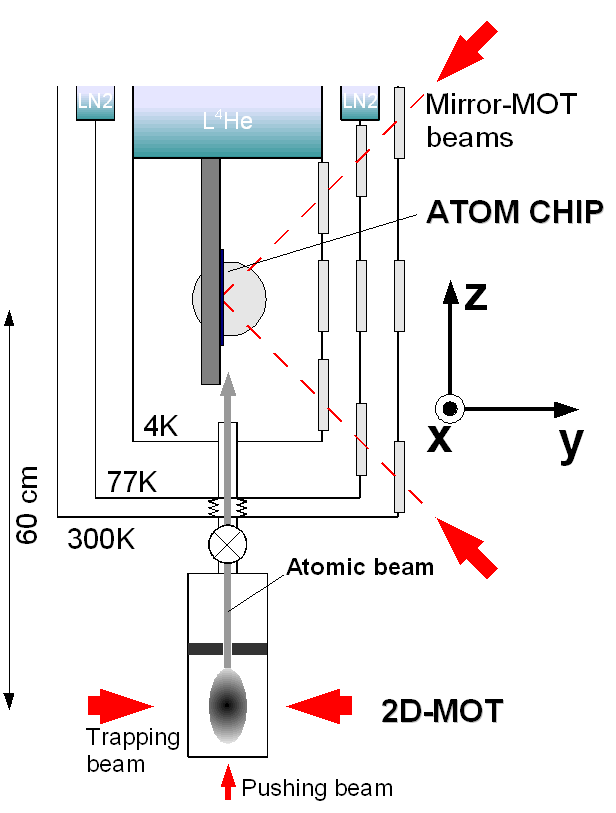}}
(b)\\
\centerline{\includegraphics[width=5.5cm]{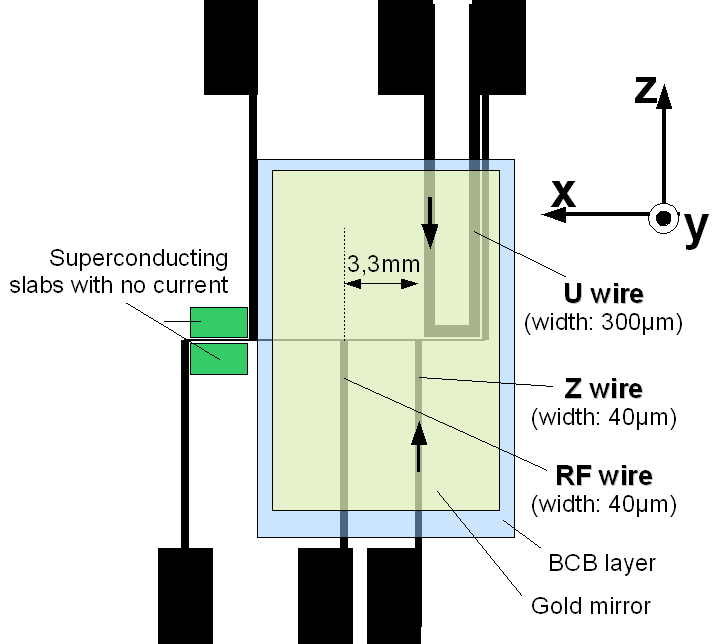}}
\caption{\label{fig:schema} (Color online) (a) Scheme of the experiment. Detailed explanations are given in the text. Laser beams along the $x$ direction are not represented. (b) Layout of the atom chip.}
\end{figure}

Our experimental chamber is shown in fig.~\ref{fig:schema}(a). Its basic operation is described
more precisely in ref.~\cite{ENS_CHIPSUPRA06}. Superconducting coils (not shown on the figure)
can create a homogeneous bias field $(B_x, B_y, B_z)$ in the vicinity of the chip. A low-velocity intense beam of $^{87}$Rb atoms
is produced in an external UHV chamber at room temperature. The atomic
flux is $1.5 \cdot 10^7$~atoms/s with a velocity distribution between 10 and
20~m/s. The atomic beam propagates upwards towards the cryogenic experimental cell, and is trapped in front of the chip in a mirror-magneto optical trap (MOT). The magnetic quadrupolar field for the mirror-MOT is produced by the superposition of a bias field $(0,0,4.4)$~Gauss and of a inhomogenous field created by a centimeter-sized elongated superconducting coil~\cite{MX_SCHMIEDMAYERQUADRUPOLETRAPCHIP04} placed 1.5~mm behind the chip. The intensity and red-shift of the trapping laser beams are 13 mW/cm$^2$ and $2.7\ \Gamma$ respectively. $\Gamma=2\pi\cdot5.9~$MHz is the natural linewidth of the $D2$ line of $^{87}$Rb. About $5 \cdot 10^7$ atoms are loaded
into the mirror-MOT in 5~s. The atomic cloud has a diameter of 2~mm and is located at a distance of
2.1~mm from the surface. Its temperature is about 300~$\mu$K. The atomic sample
is then transferred into a tighter mirror-MOT, whose gradient is produced
by an on-chip U-shaped wire made of niobium (fig.~\ref{fig:schema}(b)) together with a bias field $(0,0,1)$~Gauss. The transfer from the first macroscopic MOT to the on-chip U-MOT lasts 20~ms and its efficiency is 85~\%. In the next step, lasting also 20~ms, the bias field is linearly increased to $(0,0,4.4)$~Gauss, while the current in the U wire $I_U$ reaches 4.5~A. 
Hence, the resulting quadrupole field gradient increases from 1~Gauss/cm to 21~Gauss/cm. As a result, the atomic cloud is compressed and brought to a distance of 460~$\mu$m from the chip surface. In order to reduce atom losses during the compression and to cool down the sample, the red-shift of the trapping beams is simultaneously increased to $9.7\ \Gamma$. At the end of the procedure, the sample contains $1.2\cdot10^7$ atoms at
a temperature of $80$~$\mu$K. Its dimensions are 380~$\mu$m along the $y$ and $z$ directions and
1200~$\mu$m along the $x$ direction.

In order to increase the initial phase space density in the magnetic trap before evaporation, two additional steps are implemented with respect to the sequence described in ref.~\cite{ENS_CHIPSUPRA06}. Before switching on the magnetic Ioffe-Pritchard trap, we first perform an optical molasses stage, followed by an optical pumping in the low-field seeking state $|F=2,m_F=2\rangle$, which will be used for the trapping. 
During the optical molasses stage (duration : 2ms), the magnetic quadrupolar field of the mirror-MOT is switched off while the laser beams stay on. 
At the same time, the trapping beam intensity decreases to 65~$\mu$W/cm$^2$, and the red-shift of each beam is set to 10.2~$\Gamma$.

In the beginning of the molasses stage, the current in the U wire is set to 0~A in less than 100~$\mu$s. Due to the finite inductance of the coils, the bias field decreases more slowly. After 1~ms, its magnitude is found to be below 40~mGauss. We have checked this low value by using the atoms as a local probe for Hanle effect. After the extinction of the field, the molasses cooling is kept active during an extra period of 1~ms.  At the end of this procedure, the cloud contains $1.1\cdot10^7$ atoms at 20~$\mu$K. This temperature is about ten times smaller than the calculated depth of the magnetic trap in which the atoms will be transferred.

Next, we proceed to optical pumping into the $|F=2,m_F=2\rangle$ state. We first pump all the atoms in the hyperfine level $F=2$. The trapping beams on the $F=2 \rightarrow F'=3$ transition are turned off at the end of the molasses stage, while the repumping laser on the $F=1 \rightarrow F'=2$ transition remains on (intensity~:~30~$\mu$W/cm$^2$).
We then pump the atoms in the $m_F=2$ magnetic Zeeman sublevel. A 500~$\mu$s laser pulse is sent in presence of a $(0,2,0)$~Gauss bias field.
The optical pumping laser is frequency stabilized on the $|F=2,m_F=0\rangle \rightarrow |F'=2,m_{F'}=0\rangle$ transition and $\sigma^+$ polarized. Its intensity is 250~$\mu$W/cm$^2$. The laser is sent along the $y$ axis, and retroreflected on the chip gold mirror. No significant heating of the sample is observed. The optical pumping stage increases by a factor of 3 the population in $|F=2,m_F=2\rangle$.

The atomic sample is then transferred into the on-chip Ioffe-Pritchard trap. We first switch off all the lasers. The current in the Z wire is set to 1.4~A, and the bias field is changed from $(0,2,0)$ to $(0,0,4.4)$~Gauss in approximatively 200~$\mu$s. This fast switching time is obtained, in spite of the large inductance of the coils, with the help of a previously loaded capacitor. Once the desired value of the field is reached, the current is controlled by an external power supply, with a slower response time. During the transfer to the Ioffe-Pritchard trap, the quantization axis, determined by the total magnetic field at the cloud position, rotates from $y$ to $-x$. The atomic spins adiabatically follow this rotating field. 

Combining this fast switching procedure with the molasses cooling and optical pumping, we transfer up to $3.5(5)\cdot 10^6$ atoms in the magnetic trap (transfer efficiency from the U-MOT : 32~\%). After a 50~ms trapping time, the temperature of the atomic sample is 20~$\mu$K.

\section{Bose-Einstein condensation}

The last step towards condensation is performed by forced evaporative radiofrequency cooling ~\cite{TR_CORNELLWIEMANNOBEL02}.
In order to speed up the evaporation process, we first increase the elastic collision rate by an adiabatic compression of the cloud. The bias field is raised to $(-13,0,26)$~Gauss in 100~ms. After the compression, the Zeeman transition frequency towards untrapped states at the bottom of the trap potential is approximately 9~MHz. This value minimizes the influence of
technical noise, present at lower frequencies, at the expense of a reduced confinement of the cloud.
With these settings, the atoms are held in a ``tight trap'', located 85~$\mu$m away from the surface. Calculations of the magnetic field give axial and longitudinal trapping frequencies equal to $2\pi\times6$~kHz and $2\pi\times100$~Hz respectively.

We observe the atoms by absorption imaging after releasing the trap. The image is formed onto cooled-CCD cameras. The observation direction is either normal to the chip surface (``front" observation along $y$, magnification 7.5~$\mu$m/pixels) or along an horizontal axis at an angle of $11$~deg with respect to $x$ (``side" observation, magnification 9.2~$\mu$m/pixels). In the latter case, we observe both the direct image of the cloud and its reflection on the chip. The distance between the cloud and the surface can thus be precisely determined. The optical resolution for both directions is of the order of 1 pixel. For large time of flight and thermal atomic ensembles, the final size of the gaussian density profile directly reveals the atomic temperature. In order to avoid collisions of the cloud with the chip surface during its expansion, we first decompress adiabatically the trap in 100~ms to the final value of the bias field $(0,0,8.4)$~Gauss before switching off all the magnetic fields.  The cloud center is then located 380~$\mu$m away from the surface. The tranversal and longitudinal frequencies in this ``loose trap'' are $\omega_\perp=\omega_y=\omega_z\ = 2 \pi \times$450(100)~Hz and  $\omega_\parallel=\omega_x = 2 \pi \times$30(10)~Hz respectively. All the temperatures and number of atoms given in the following are measured in the loose trap. 

\begin{figure}
(a)\\
\centerline{\includegraphics[width=7cm]{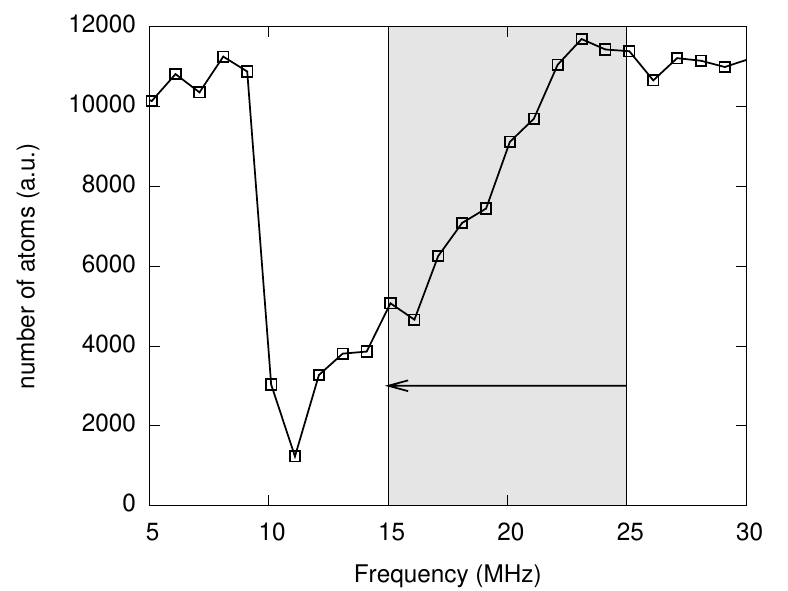}}
(b)\\
\centerline{\includegraphics[width=7cm]{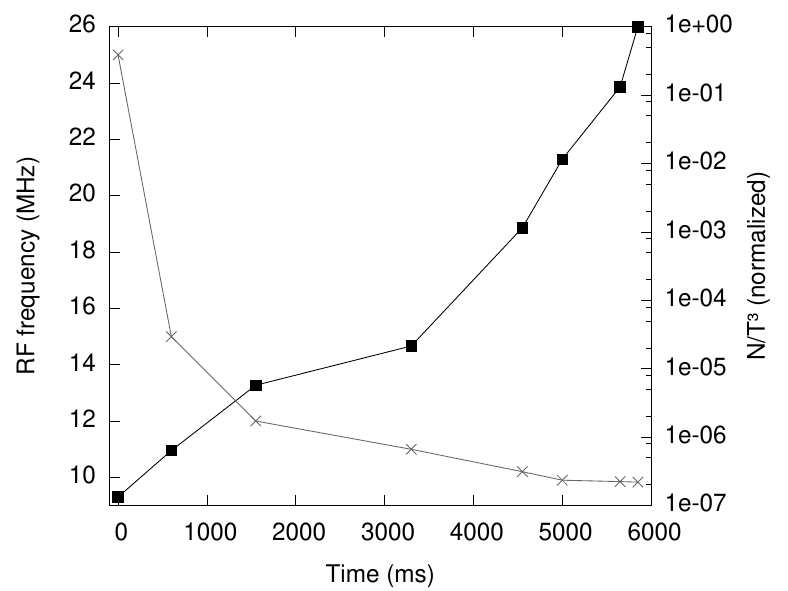}}
\caption{(a) Number of atoms remaining in the trap as a function of the frequency of the RF field exciting Zeeman transitions towards untrapped states. The latter is applied while the atoms are in the tight trap. It gives a rough idea of the energy distribution of the cloud. (b) Time evolution of the RF knife frequency during the evaporation process ($\times$). The  $N/T^3$ ratio at the end of each segment is displayed on the same graph~($\blacksquare$). It is normalized to 1 at the end of the process.}
\label{fig:spectroRF}
\end{figure}

Forced evaporative cooling of the sample is performed in the tight trap, in which the elastic collision rate is high. We ramp down a RF knife whose initial frequency corresponds to the energy of the hottest atoms in the trap. The RF knife is produced with the help of a current that is fed in an extra superconducting wire, located on the chip about 3.3~mm away from the atoms (cf fig.~\ref{fig:schema}(b)). In these conditions, elastic collisions produce hot atoms which are expelled, while the mean temperature of the remaining sample decreases. The frequency of the knife is ramped down in order to match the temperature decrease, and thus force the process. Our RF ramp is made of seven linear parts, each of them being optimized as described in the following.

For the first linear part, three parameters have to be chosen: the total duration of the segment and its initial and final frequencies. The latter can be
determined using Zeeman spectroscopy on the trapped cloud. Fig.~\ref{fig:spectroRF}(a)  shows the number of atoms still trapped after having been held for 100~ms in the tight trap. During that period, the RF knife current is sent at a fixed frequency in the setup. The resulting RF magnetic field induces Zeeman transitions for the resonant atoms. Before the first segment of the ramp, if the frequency is below 9~MHz or above 25~MHz, no atom is resonant with the field, and no atomic losses are observed. If the frequency lies between these two values, a significant part of the atoms is ejected from the trap. This measurement provides an estimation of the atomic energy distribution and allows us to choose the frequencies of the first step of the evaporative cooling. The frequency of the first segment is linearly decreased from 25~MHz, where no atoms are lost, down to 15~MHz, which corresponds to the half width of the frequency distribution of fig.~\ref{fig:spectroRF}(a) (gray area). In order to optimize the duration of this first linear segment, we measure the remaining number of atoms $N$ and the final temperature of the cloud $T$ in the loose trap for different ramp durations. We then compute $N/T^3$, which is proportional to the phase space density in an ideal harmonic trap. The optimum duration of the first segment has thus been determined to be 600~ms, which is the value that maximizes $N/T^3$.

\begin{figure}
\centerline{\includegraphics[width=7.2cm]{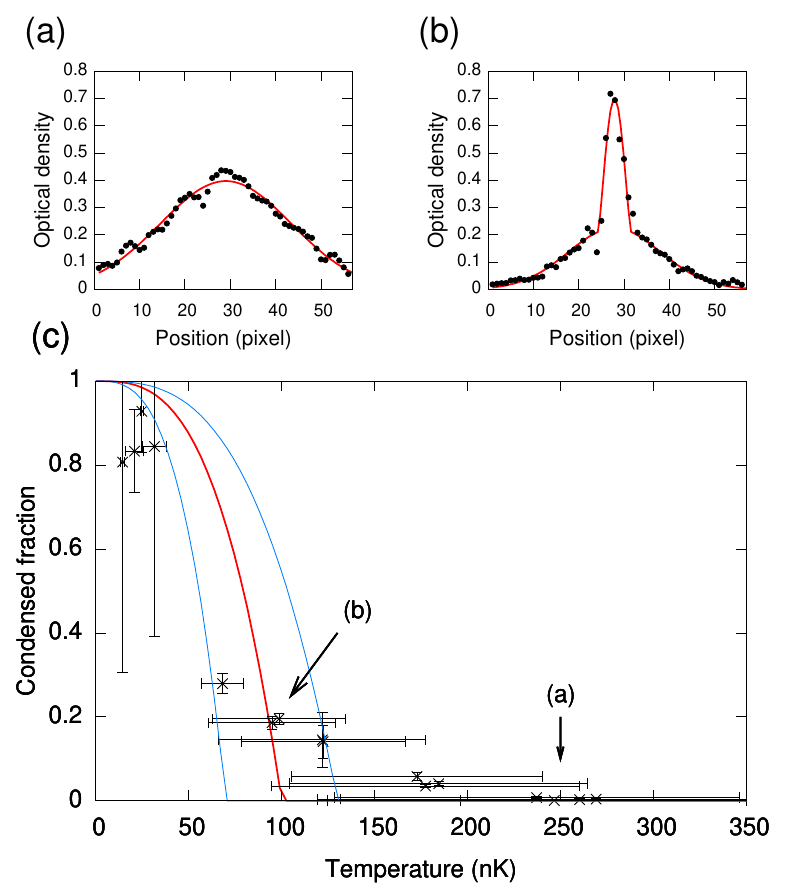}}
\caption{(Color online) (a) Horizontal cross-section of the front absorption image after 17~ms of expansion, for a final evaporation ramp frequency $f=$9.871~MHz. The solid line is a gaussian fit. (b) Same as (a) with $f=$9.841~MHz. It shows the emergence of a condensed fraction, fitted by a Thomas-Fermi distribution superimposed on a thermal background (solid line) (c) Condensed fraction of atoms as a function the cloud temperature $T$. The latter is varied  by adjusting the final frequency of the evaporation ramp. The solid lines are theoretical fractions in absence of interactions for the transition temperatures $T_c=$70, 100 and 130~nK}
\label{fig:condensedfraction}
\end{figure}

Our complete evaporation ramp finally ranges from 25~MHz down to an adjustable frequency between 9.811~MHz and 9.881~MHz. Fig.~\ref{fig:spectroRF}(b) shows the evolution of the RF knife frequency during the process, as well as the $N/T^3$ ratio, which increases by more than 7 orders of magnitude. The latter has been normalized to 1 at the end of the ramp, where $f=$9.841~MHz. The final frequency of the last segment is adjusted with a resolution of 5~kHz in order to have a good control of the temperature. Fig.~\ref{fig:condensedfraction}(a) and (b) show two cross-sections in the absorption image of the atomic cloud for the final frequencies $f=$9.871~MHz and $f=$9.841~MHz respectively. Both images are taken along the front observation direction, 17~ms after releasing the cloud from the loose trap. A sharp peak clearly appears on fig.~\ref{fig:condensedfraction}(b) on top of a broader thermal gaussian profile. The bimodal distribution is a strong evidence for the presence of a BEC. The condensed part of the sample can be fitted by a 2D Thomas-Fermi distribution, whose amplitude is used to determine the fraction of condensed atoms for fig.~\ref{fig:condensedfraction}(c). The cloud temperature $T$, in abscissa of the graph, has been obtained by fitting the width of the thermal fraction with a gaussian. However, the thermal fraction is too small to yield a precise fit in the case of the first four points of the graph (lowest temperatures). In those particular cases, $T$ is extrapolated from a linear fit of its dependency as a function of the final frequency of the ramp. The solid lines in fig.~\ref{fig:condensedfraction}(c) correspond to the fit of the condensed fraction in absence of interaction between atoms for three different critical temperatures ($T_c=$70, 100 and 130~nK), which are qualitatively compatible with the experimental points. The corresponding critical temperature and number of atoms at the onset of BEC are $T_c=100(30)$~nK and $N=1.0(5)\cdot10^4$ respectively. For this number of atoms and the trapping frequencies obtained from numerical calculations, the expected theoretical critical temperature in the loose trap is $175(75)$~nK~\cite{TXT_STRINGARI}.

\begin{figure}
\centerline{\includegraphics[width=7cm]{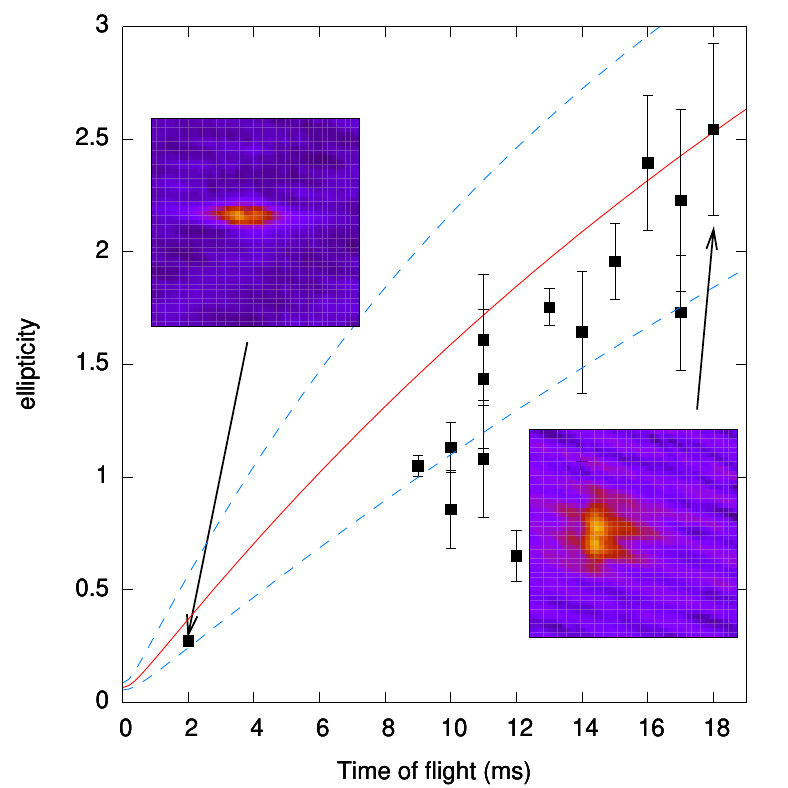}}
\caption{(Color online) Ellipticity of the atomic cloud during the time of flight. The insets show two images of the cloud at the begining and the end of the expansion. In the latter case, the observed fringes are due to interference effects between successive cryogenic viewports. The solid line is the theoretical predictions with no adjusted parameter. The two dashed lines take into account the uncertainties on the trapping frequencies.}
\label{fig:ellipticity}
\end{figure}

Further evidence for Bose-Einstein condensation is given by the inversion of ellipticity of the condensed part during a time of flight expansion. From the absorption images, we can extract the widths of the Thomas-Fermi profiles along the longitudinal ($w_x$) and transverse ($w_z$) directions. The ellipticity of the cloud is defined as the ratio $w_z/w_x$. Fig.~\ref{fig:ellipticity} shows the ellipticity as a function of the time of flight after releasing the atoms. It clearly displays an inversion between the $x$ and $z$ axes of the cloud. The solid line is provided by the scaling law of ref.~\cite{TR_CASTINDUMEXPANSION96} with no adjustable parameter. It qualitatively fits the experimental data. The two dashed lines account for the uncertainties on the trapping frequencies.

\section{Conclusion and perspectives} We have prepared the first BEC on a cryogenic superconducting atom-chip. The number of condensed atoms and the temperature are compatible with earlier results obtained in atom chips experiments. These results are a good starting point to lifetime measurements in the vicinity of superconductors. For this purpose, we will move the atoms away from the gold mirror, towards a  region of the chip made only of superconducting material (fig.~\ref{fig:schema}(b)). This could be achieved using a magnetic conveyor belt, as in ref.~\cite{TR_HANSCHCONVEYOR01}. We also plan to excite the very dense atomic cloud towards Rydberg states. At the end of the evaporation ramp in the tight trap, the cloud is contained within a radius of a
few $\mu$m. This situation is appropriate for strong dipole-blockade~\cite{TR_EYLERBLOCKADE04, TR_WEIDEMULLERDENSE04, TR_PILLETBLOCKADE06, TR_PFAUBLOCKADE07}, possibly leading to the preparation of a single Rydberg atom~\cite{TR_SAFFMANN02} within the condensate. This Rydberg atom could be further manipulated in an electrodynamical on-chip trap~\cite{ENS_RYDBERGTRAP04, ENS_EPJDTRAP}.

\acknowledgments
Laboratoire Kastler Brossel is a laboratory of ENS and Universit\'{e} Pierre et
Marie Curie, associated to CNRS (UMR 8552). We acknowledge support of the
European Union (CONQUEST and SCALA projects, Marie Curie program), of the Japan
Science and Technology corporation (International Cooperative Research Project~:
``Quantum Entanglement''), and of the R\'egion Ile de France (IFRAF and C'nano
IdF institutes). C.R. acknowledges support from DGA.



\end{document}